\documentclass[aps,pre,floats,floatfix,showpacs,unsortedaddress,balancelastpage,preprint]{revtex4-1}
\usepackage[dvipdf]{graphicx}
\usepackage{amsmath}
\usepackage{amsfonts}
\usepackage{mathrsfs}
\usepackage[usenames]{color}

\newcommand{\dd}{\,\mathrm{d}}

\newcommand{\thalf}{\tfrac{1}{2}}
\newcommand{\pd}{\partial}

\newcommand{\Pd}[3][{}]{\frac{\pd^{#1} #2}{\pd #3^{#1}}}
\renewcommand{\vec}[1]{\mathbf{#1}}
\newcommand{\sgn}{\mathop{\rm sgn}}

\begin{document}

\title{Topological and geometric decomposition of nematic textures}
\author{Simon \v{C}opar}
\affiliation{Faculty of Mathematics and Physics, University of Ljubljana, Jadranska 19, 1000 Ljubljana, Slovenia}
\author{Slobodan \v{Z}umer}
\affiliation{Faculty of Mathematics and Physics, University of Ljubljana, Jadranska 19, 1000 Ljubljana, Slovenia}
\affiliation{Jo\v{z}ef Stefan Institute, Jamova 39, 1000 Ljubljana, Slovenia}
\date{\today}


\begin{abstract}
  Directional media, such as nematic liquid crystals and ferromagnets,
  are characterized by their topologically stabilized defects in
  directional order. In nematics, boundary conditions and
  surface-treated inclusions often create complex structures, which
  are difficult to classify. Topological charge of point defects in
  nematics has ambiguously defined sign and its additivity cannot be
  ensured when defects are observed separately. We demonstrate how the
  topological charge of complex defect structures can be determined by
  identifying and counting parts of the texture that satisfy simple
  geometric rules.  We introduce a parameter called the defect rank
  and show that it corresponds to what is intuitively perceived as a
  point charge based on the properties of the director field. Finally,
  we discuss the role of free energy constraints in validity of the
  classification with the defect rank.
\end{abstract}

\maketitle

\section{Introduction}

Various states of matter we know today are characterized by
interactions and symmetry of their microscopic building blocks. Local
order is reflected in mesoscopic quantities, such as the phase of
electromagnetic waves, crystalline order, nematic director,
magnetization, dielectric tensor and others. We call these quantities
order parameters. A prominent example of partially ordered matter are
liquid crystals, which are characterized by orientational order and
liquid-like positional freedom of the molecules. Nematic liquid
crystals exhibit topologically prescribed defects, which in
combination with surface-treated colloidal particles \cite{musevic},
nanoparticles or general confinement \cite{belini,serra}, enable
creation of complex materials with tunable optical response
\cite{humar}.

The nematic order can be represented with an order parameter tensor
that contains both the magnitude of the ordering and the directional
information \cite{degennes}. In studies of defect topology, the
principal eigenvector of this tensor, the director, is a common choice
to represent the average direction of molecular alignment. Vectors
with unit size lie on a sphere ($S^2$), which is a topologically
nontrivial space. Consequently, boundary conditions can exist that
cannot be smoothly interpolated in bulk without introducing defects
\cite{mermin,trebinrev}. In contrast to Heisenberg ferromagnets, which
have the ground state manifold with exact topology of a sphere,
nematic liquid crystals have an additional rule that opposite vectors
represent the same state, as the molecules have no back-front
distinction \cite{volomin}. One consequence of this property is that
beside the point defects, nematics also allow line defects. Beside
that, the sign of the vector field representing the director is
ambiguous and therefore the same holds for the invariants of the field
which are odd in the director, such as the topological charge of point
defects \cite{mermin}.

The growing interest for controlled formation of complex defect
structures in constrained nematic liquid crystals
\cite{musevic,belini,serra} is increasing the demand for visualization
and simple classification of complex nematic director fields.  This
stimulated us to derive a natural expression for the topological
charge that does not involve integration and reflects its discrete
nature. We accomplish this by decomposing the director field into a
hierarchy of geometric primitives, which enables intuitive
understanding of the topological charge based on the qualitative
features of the director field. We extend the classification from
strictly topological to a more general one that has the advantage of
differentiating topologically equivalent states with different
energies. We introduce a quantity called the defect rank and draw
parallels with the topological charge and naming conventions for point
defects.

\section{Patches, boundaries and grains}

If the vector heads are artificially prescribed, the director field
$\vec{n}$ becomes a proper vector field. Volumes of interest, such as
those that include defects, can then be classified using the
topological charge, which enumerates the defects with the elements of
the second homotopic group \cite{mermin,volomin}. As topological
properties are invariant to the specific shape of the surface that
encloses the test volume, a spherical volume is assumed from here on
without the loss of generality. For a sphere, parametrized by the
spherical angles $\theta$ and $\phi$, the Gauss integral for the
topological charge \cite{klemanrev,klem,trebinrev,kotiuga} reduces to
the following expression,
\begin{equation}
  q=\frac{1}{4\pi}\iint \vec{n}\left(\Pd{\vec{n}}{\theta}\times\Pd{\vec{n}}{\phi}\right)\dd \theta \dd \phi.
  \label{eq:qdef}
\end{equation}

\begin{figure}
  \includegraphics[width=\textwidth]{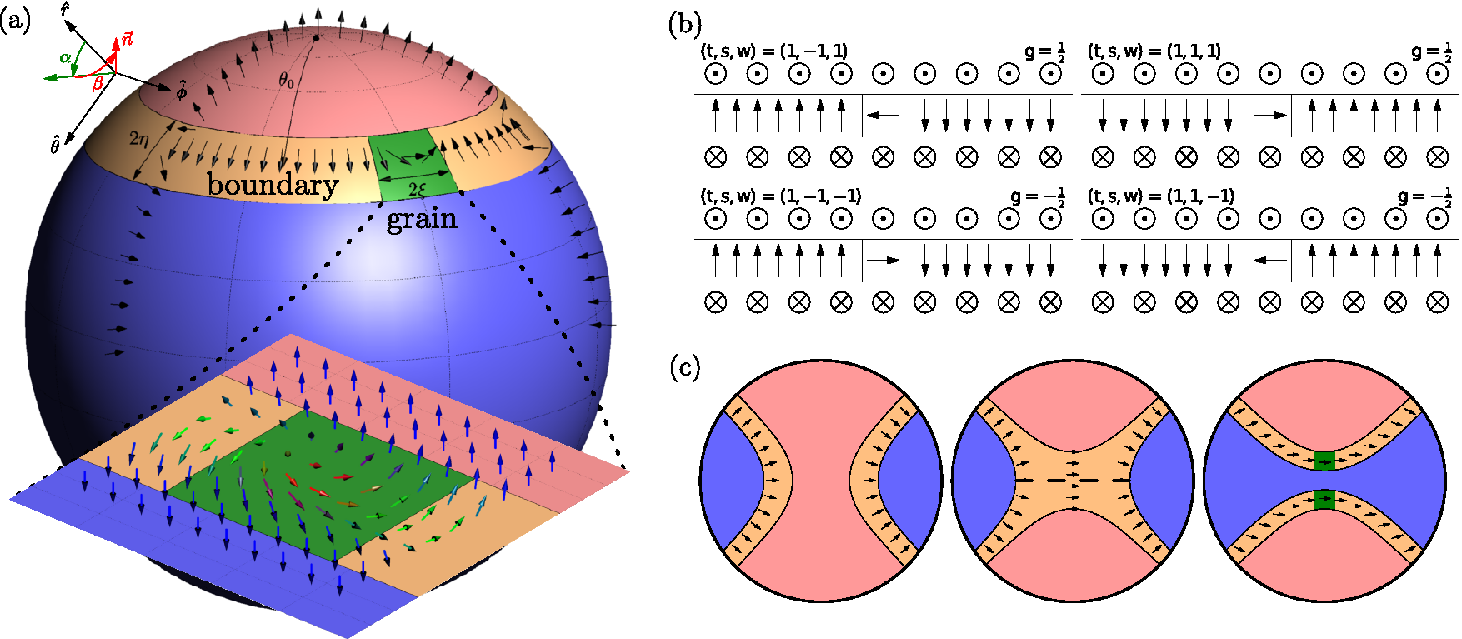}
  \caption{\label{fig:fig1}
    Geometric decomposition of a surface texture on a sphere.
    {(a)} Any texture can be smoothly deformed to make the director
    perpendicular to the sphere everywhere except in narrow
    transition zones. This decomposes the surface into patches with
    perpendicular director, boundary loops with parallel director
    and point-like grains with the director rotating in-plane for
    half a turn. The inset shows the continuous texture of a grain.
    {(b)} Grains contribute $g=\pm1/2$ to the topological charge,
    depending on the sense of the director rotation. The table
    shows four possibilities, together with corresponding $(t,s,w)$
    parameters that define the director's orientation in patches,
    boundaries and grains, respectively.
    {(c)} Merging and dividing of patches by continuous transformations
    preserves the topology of the texture. The change in the number of
    patches is compensated by spontaneous creation of grains.
  }
\end{figure}

This expression involves integration over all the details of the
director field $\vec{n}$, which can be eliminated, if the director
$\vec{n}$ on the enclosing sphere (the surface texture) is decomposed
into geometric primitives, over which the integration can be performed
in advance. According to the component of $\vec{n}$ along the surface
normal, the sphere can be split into patches of outgoing vector field
and patches of ingoing vector field. These patches are separated by a
disjoint set of closed loops (boundaries) where the director lies
tangentially to the enclosing sphere. On the boundaries, the director
still has one degree of freedom -- its direction in the tangent
plane. With the same approach as for the patches, the boundary can be
split into parts with a constant orientation along the normal to the
boundary curve. These are separated by points (grains), where
$\vec{n}$ is directed tangentially along the boundary
(Fig.~\ref{fig:fig1}a). Degenerate cases, such as textures with
touching boundaries or wide areas where director is parallel to the
surface, can always be regularized by an infinitesimal perturbation of
the director field.

The parametrization of the sphere with the polar angle $\theta$ and
the azimuth $\phi$ generates a local coordinate frame
$(\hat{r},\hat{\theta},\hat{\phi})$, where $\hat{\theta}$ and
$\hat{\phi}$ lie parallel to the surface
(Fig.~\ref{fig:fig1}a). Within this coordinate frame, we introduce
local spherical coordinates with angles $\alpha$ and $\beta$ to
express the director $\vec{n}$,
\begin{equation}
  \vec{n}=t(
  \hat{r}\cos\alpha+\hat{\theta}\sin\alpha\cos\beta+\hat{\phi}\sin\alpha\sin\beta
  )
  \label{eq:npar}
\end{equation}
where $t=\pm 1$ simplifies the calculation for director fields with
reversed vectors. Without the loss of generality, each angle can vary
in only one of the directions: $\alpha=\alpha(\theta)$ and
$\beta=\beta(\phi)$.

To calculate the contributions of patches, boundaries and grains to
the topological charge (Eq.~\ref{eq:qdef}), a model for
$\alpha(\theta)$ and $\beta(\phi)$ is needed. On every patch, the
director $\vec{n}$ can be combed perpendicularly to the surface
without changing the topology, because the condition for the normal
component to have a constant sign locally restricts the ground states
to a half-sphere, which is topologically trivial (contractible to a
point). This is achieved by setting $\alpha=\beta=0$ so that the
director points directly outwards if $t=1$ and inwards if $t=-1$.  For
a single patch, shaped like a spherical cap bounded by a circle of
constant latitude at $\theta=\theta_0$ (Fig.~\ref{fig:fig1}a), the
contribution to the topological charge follows,
\begin{equation}
  q_p=\frac{1}{4\pi}\int_0^{2\pi}\int_0^{\theta_0} \vec{n}\left(\Pd{\vec{n}}{\theta}\times\Pd{\vec{n}}{\phi}\right)\dd \theta \dd \phi
  =\frac{1}{2}t(1-\cos\theta_0).
  \label{eq:qp}
\end{equation}
This contribution is proportional to the fraction of the sphere
covered by the patch and its sign depends on whether the director
points inwards or outwards from the surface, which is given by the
signature $t=\pm 1$.

On the boundary, the director makes a smooth turn from outward to
inward direction. A boundary without grains is modeled by setting
$\beta=0$ and choosing $\alpha$ to do a half-turn within a narrow
transition zone between the boundary and the patch,
\begin{equation}
  \alpha(\theta)=s\frac{\pi}{2}\left(1+\frac{\theta-\theta_0}{\eta}\right).
  \label{eq:alfass}
\end{equation}
$\eta$ is the half-width of the transition zone around the boundary,
and the boundary signature $s=\pm 1$ selects the in-plane orientation
of the director at the boundary, $\vec{n}=st\hat{\theta}$
(Fig.~\ref{fig:fig1}a). To simplify the calculation, the width $\eta$
of the transition zone is limited to zero,
\begin{equation}
  q_b=\lim_{\eta\to 0}\frac{1}{4\pi}\int_0^{2\pi}\int_{\theta_0-\eta}^{\theta_0+\eta} \vec{n}\left(\Pd{\vec{n}}{\theta}\times\Pd{\vec{n}}{\phi}\right)\dd \theta \dd \phi
  =t\cos\theta_0.
\label{eq:qb}
\end{equation}
The result does not depend on $s$, therefore the orientation of the
director on the boundary does not influence the topological charge.

The contribution of a single grain is obtained in a similar fashion.
The reverse of the in-plane orientation of the director is smoothly
interpolated by $\beta(\phi)=w \pi/2(1+(\phi-\phi_0)/\xi)$, with $\xi$
being the half width of the transition in $\hat{\phi}$ direction
(Fig~\ref{fig:fig1}a,inset). Parameter $w=\pm 1$ governs in which way
the director points at the grain, $\vec{n}=(tsw)\hat{\phi}$. A limit
$\xi\to 0$ yields the contribution of the grain.
\begin{equation}
  g=\lim_{\eta,\xi\to 0}\frac{1}{4\pi}\int_{\phi_0-\xi}^{\phi_0+\xi}\int_{\theta_0-\eta}^{\theta_0+\eta} \vec{n}\left(\Pd{\vec{n}}{\theta}\times\Pd{\vec{n}}{\phi}\right)\dd \theta \dd \phi
  =\frac{tw}{2}
\label{eq:qg}
\end{equation}

\begin{figure}
  \includegraphics[width=0.5\textwidth]{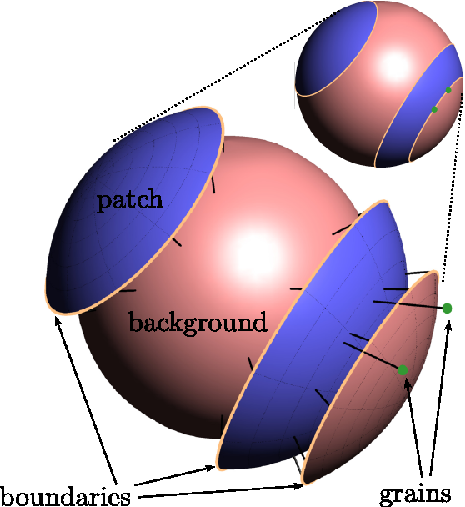}
  \caption{\label{fig:fig2}
    Boundaries divide the texture into patches with arbitrary
    positioning, shape and hierarchical nesting order. If the texture
    is instead viewed as a superposition of layers, the boundaries
    can be counted together with patches. In such decomposition, each
    patch contributes $\pm1$ to the topological charge, depending on
    whether its direction is oriented inwards or outwards, here
    represented with alternating colors.
  }
\end{figure}

Consider a texture, divided by $N$ boundaries into $N+1$ patches of
arbitrary shapes. Every boundary can have an even number of grains,
since the continuity requires an even number of director flips on a
closed circuit. The topological charge $q$ of the entire texture then
splits into a sum of elementary contributions
(Eqs.~(\ref{eq:qp},\ref{eq:qb},\ref{eq:qg})),
\begin{equation}
q=\sum_{i=0}^N q_{pi}+\sum_{i=1}^N q_{bi}+\sum_{i}^{M} g_i.
\end{equation}
Instead of adding patch contributions $q_{pi}$, which are proportional
to the areas of the patches, multiplied by the signature $t$
(Eq.~\ref{eq:qp}), we start with a uniform background sphere and
progressively build the desired configuration. For convenience, we
choose the background with the director pointing outwards, $t=1$. To
recover the desired configuration, $N$ patches must be added. Each
patch contribution is counted twice, once to cancel out the background
and once to account for its own contribution. This way, all the
patches are shaped as spherical caps (Fig.~\ref{fig:fig2}) and their
contributions can be grouped with their respective boundaries,
\begin{equation}
\sum_{i=0}^N q_{pi}+\sum_{i=1}^N q_{bi}=q_0+\sum_{i=1}^N(2 q_{pi}+q_{bi})=1+\sum_{i=1}^N t_i.
\end{equation}
The background patch $q_0$ has the topological charge of a radial
hedgehog, $q_0=\chi/2=1$, where $\chi$ is the Euler characteristic of
the surface \cite{geom1,lavworlds}. The dependence on the polar angle
$\theta$ has disappeared, since the shape and position of the
boundaries are not topologically relevant. Any number of boundaries
can be nested one inside another, which results in patches with
alternating $t$ being added to the sphere (Fig.~\ref{fig:fig2}).

The grain contribution $g=tw/2$ can be associated with geometric
features of the texture. The director $\vec{n}$ rotates around the
surface normal as we move along the boundary. There are two possible
directions in which we can trace the boundary, but they are
invariantly distinguishable, as there are patches of opposite
signatures on each side. For a choice of the direction that has the
positive signature patch on its left side, the grain contribution is
$+1/2$ if the director rotates anticlockwise and $-1/2$ otherwise
(Fig.~\ref{fig:fig1}b).

The charge decomposition in the form
\begin{equation}
q=1+\sum_i^N t_i+\sum_i^M g_i
\label{eq:qfinal}
\end{equation}
applies not only to the nematic director, but to any medium that can
be described with a unit vector order parameter. It works for
spherical enclosing surfaces, with possible generalization to surfaces
of arbitrary genus.  We assumed the vector field on the enclosing
surface is non-singular. Singular objects, like boojums and other
surface defects, contribute additional terms to the expression
\cite{kurik,volovik,lavworlds}.

Patches, boundaries and grains are two-, one- and zero-dimensional
geometric features of the surface texture. This decomposition is
similar to the notion of domain walls and defect lines in
ferromagnetic systems \cite{walls,nabarro}, except in our case we
observe 2D texture on a closed surface instead of bulk 3D field. Our
boundaries are similar to cross sections of magnetic N\'eel walls and
the grains correspond to cross sections of Bloch lines. However, in
magnetic systems, the domain walls and lines naturally prefer narrow
transition zones \cite{walls}, whereas in nematics narrow transition
zones are merely a computational tool.

\section{Free energy constraints}

One of the reasons for calculating the topological invariants is to
detect similarity between director fields. Take two surfaces that
enclose closed volumes in the bulk. If we cut out the volumes and
exchange them, we can only reconnect them smoothly with the ambient
director field if they are topologically equivalent. Topological
charge tests for compatibility of bulk volumes of the medium judging
only by the director on their surfaces, so the charge can be assigned
to the surface texture itself. Compatibility of two surface textures
neither implies additivity of the topological charge hidden inside nor
the invariance to the choice of the enclosing surface. We are free to
add further restrictions and thus refine the classification.

Despite the topological equivalence of two volumes, the transition
between them may not be possible due to the energy constraints, which
depend on the medium and the choice of the enclosing surfaces. The
structures can be deemed equivalent if one can be transformed smoothly
into the other without deviating significantly from the local free
energy minimum (if there is a transformation that avoids major free
energy barriers). The deformation part of the free energy is related
to the spatial derivatives of the director. A general transformation
can both deform the shape of the medium and independently rotate the
director locally. However, if the director is treated like it is
pinned to the underlying coordinate space and rotates with it, the
neighborhood of each point stays approximately the same, except for
the anisotropic distortions, which can be considered small. By
allowing only the subset of all smooth deformations that obey this
rule, topologically equivalent states with different energies can be
differentiated. In topology, such mappings are called push-forward
mappings. This type of restriction is more natural for solids, where
the underlying space corresponds to the crystal lattice. We can argue
that two structures are similar if we could deform one into the other
if they were elastic solids.

The surface texture decomposition involves expressing the director in
a coordinate frame aligned to the surface (Eq.~\ref{eq:npar}). Under a
mapping that satisfies the restriction to push-forward transformations,
the local coordinate system retains its relative position to the
director and the decomposition does not change. The texture
decomposition is therefore a property that classifies the surface
textures into classes, where the representatives of the same class can
be stitched together without significant change in free energy. The
topological charge, which can be computed from the decomposition
(Eq.~\ref{eq:qfinal}), is only part of the information included in the
full decomposition. Other invariants, like the relative position of
the boundaries and their boundary signatures $s=\pm1$, which have no
effect on the topology, are also good for classification of the
qualitative shape of the texture. In the following section we show
that these quasi-invariants are relevant for nematics, where
traditional topological charge is problematic due to the additional
symmetries of the director.

\section{Defect rank of nematic textures}

Topological charge (Eq.~\ref{eq:qdef}) has an odd parity under
inversion of the director $\vec{n}\mapsto-\vec{n}$. The additivity of
the topological charge, which holds for true unit vector fields, may
be violated because nematics allow line defects with half-integer
winding number. These defects cannot be represented as a vector field
without explicitly introducing a branch discontinuity where the vectors
flip sign. In the presence of line defects, the topological charge of
each individual point defect can be flipped just by bringing it across
a branch discontinuity, without changing its physical shape
\cite{trebinrev,mermin}. The topological charge of a cluster of point
defects cannot be determined from the individually measured charges of
constituent point defects. Instead, only even or odd parity of the
topological charge is properly defined and conserved
\cite{mermin,janich}.

Despite ambiguity, the signs of topological charge are usually
assigned by convention. The charge $+1$ is assigned to all radial
hedgehogs and $-1$ to hyperbolic hedgehogs, regardless the orientation
of the arrows on vectors, even though there is a 3D rotation in
director space that can transform one into the other
\cite{terent,kurik}. In case the hedgehogs interact in the same
medium, the continuity of the director field prescribes their relative
signs of topological charge that may clash with this convention. The
transformation from a hyperbolic to radial hedgehog involves a
rotation of the director without simultaneous rotation of the
underlying space. Such transformation does not satisfy our condition
to push-forward transformations and the incompatibility of the surface
textures means they cannot exist in the same surrounding director
field without costly elastic deformations. The transition between the
hedgehogs is thus restricted by an energy barrier, except in cases
where the experiment is specifically designed to support the
transition \cite{terent}.

\begin{figure}
  \includegraphics[width=0.5\textwidth]{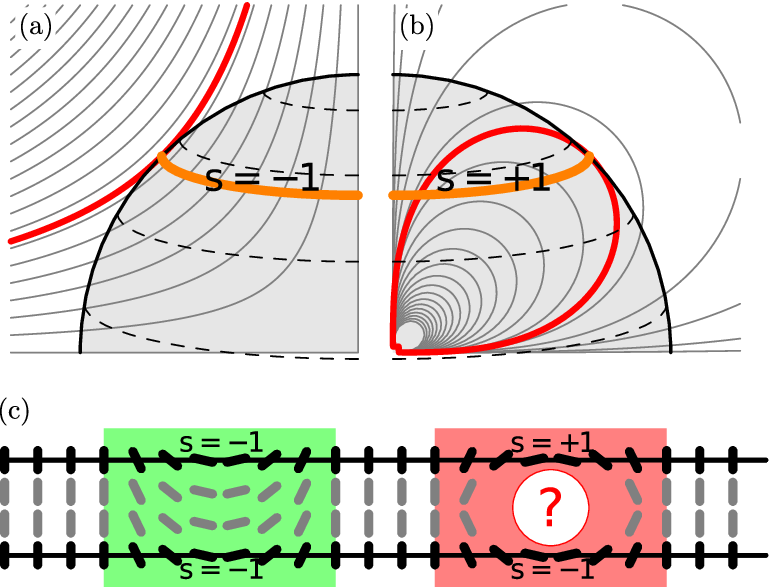}
  \caption{\label{fig:fig3}
    The geometric meaning of the boundary signature $s$.
    {(a,b)} The boundaries are the curves where the streamlines of the
    director field touch the enclosing surface. The boundary signature
    $s$ distinguishes whether the critical streamline (thicker line,
    red online) lies inside or outside the enclosing surface.
    It is not a topological invariant, but is invariant to the
    director sign reversal. The streamlines show the director in
    a radial cross section of the sphere.
    {(c)} Schematic representation of two surfaces with nematic
    textures in the cross section. Surface textures are compatible if
    their boundary signatures $s$ match. The boundaries of opposite
    signatures cannot be interpolated smoothly, unless the director is
    rotated through the third dimension while the surface stays
    undeformed, which is forbidden in our classification rules.
  }
\end{figure}

We seek a geometric invariant that matches the conventional
understanding of the hedgehog charge and preserves even or odd parity,
without introducing any inconsistencies not already present in the
topological definition. In the following section, we define the defect
rank of a point defect, $q^\ast$, and compare it to the topological
charge $q$.

The definition of the defect rank $q^\ast$ can be made consistent with
the conventions if we find a matching integer expression involving the
surface texture properties. Earlier (Eq.~\ref{eq:alfass}) we
introduced the boundary signature $s=\pm 1$ that describes how the
director behaves on the boundary. The boundary signature $s=-1$
corresponds to the director stream lines that touch the boundary from
the outside of the sphere (Fig.~\ref{fig:fig3}a) and $s=1$ to the
streamlines that touch the boundary from the inside
(Fig.~\ref{fig:fig3}b). This behavior is invariant to the inversion of
vectors, therefore $s$ is a good parameter for classification of
nematic textures. Whether the streamlines that go through the boundary
curve lie inside or outside the enclosing sphere can be tested by
observing how the normal component of $\vec{n}$ along the direction of
$\vec{n}$. A local expression $s=-\sgn((\vec{n}\cdot
\nabla)(\vec{n}\cdot \boldsymbol{\nu}))$ can be constructed, which
assigns the boundary signature to each point of the boundary and is
also invariant to the inversion of $\vec{n}$.

Two surface textures are compatible if all the boundaries and grains
are similar and arranged in a similar hierarchical pattern. To extract
an invariant that resembles the point charge, we disregard most of the
information and only focus on the boundary signatures $s$. Grains are
the points on the boundary where $s$ changes sign. The boundary
signature $s$ can be assigned to a boundary as a whole only if there
are no grains present on the texture, otherwise it is a local property
that is only constant between two grains. To find a suitable
expression for the defect rank, the classification must be restricted
to textures without grains. Two such textures match if their
boundaries have the same signature $s$, otherwise the stitching
involves rotation of the director relative to the local coordinate
frame (Fig.~\ref{fig:fig3}c). The topological charge
(Eq.~\ref{eq:qfinal}) alternates between even and odd with the number
of boundaries. This property is replicated by the defect rank, if we
define it as
\begin{equation}
  q^\ast=1+\sum_i^N s_i.
  \label{eq:tqdef}
\end{equation}
If the surface has grains in its texture, the defect rank $q^\ast$ is
not defined. The textures with grains can be understood as
intermediate states through which the texture must pass in order to
change its defect rank. For the defect rank to be a useful parameter,
such intermediate states must have a higher energy, although the
energy constraints are not absolutely prohibiting, like topological
are. If the energy barrier is removed, which can be accomplished by
external fields, boundary frustration or heating, the transition
occurs and the texture changes \cite{terent,kurik}.  The entire
texture on the enclosing surface can still be used for more precise
comparison and testing for compatibility. The texture decomposition
can be seen as a generalization of the traditional topological charge,
as the latter is included in the former, as given by
(Eq.~\ref{eq:qfinal}).

\begin{figure}
  \includegraphics[width=0.5\textwidth]{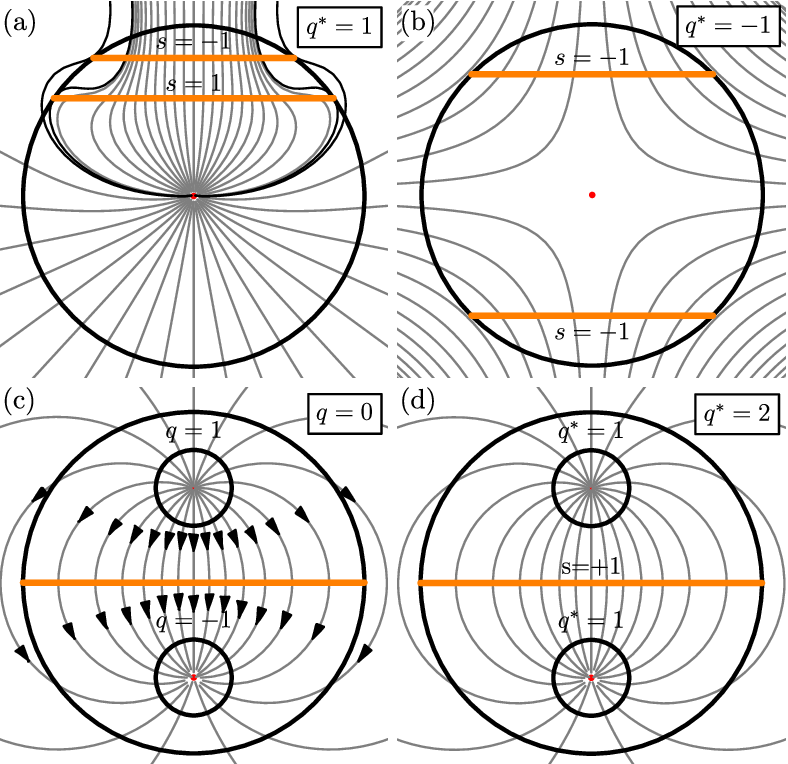}
  \caption{\label{fig:fig4}
    Point defect charge in nematics visualized with director field streamlines in cross section.
    {(a)} The signatures of boundary pairs, caused by director field
    bending are opposite and conserve the defect rank.
    {(b)} Hyperbolic hedgehog is identified by the defect rank of
    $q^\ast=-1$.
    {(c)} For the dipole configuration, strict topological
    definition assigns the opposite charges to the two hedgehogs,
    corresponding to the direction of the arrows and assigns the
    charge $q=0$ to the whole structure.
    {(d)} The defect rank treats the hedgehogs equally and
    amounts to $q^\ast=2$, the same as expected
    with na\"ive counting of the hedgehogs. The arrows are not needed
    for determination of the defect rank.
  }
\end{figure}

The boundary signature $s$ is not a topological invariant, but
nevertheless has interesting properties which resemble conservation
laws needed for effective classification. Two neighboring boundaries
with opposite signatures can annihilate, as they represent a ``fold''
in the director field (Fig.~\ref{fig:fig4}a). The definition of
$q^\ast$ matches the convention for the radial hedgehog, which has
no boundaries, ($q^\ast=1$), and hyperbolic hedgehog that has two
boundaries with negative signature ($q^\ast=1+2\cdot(-1)$,
Fig.~\ref{fig:fig4}b). The singular case when the boundary is shrunk
to a point corresponds to a point defect sitting on the enclosing
surface \cite{volovik}. Pushing additional point defect with rank $s$
inside the enclosing surface adds a boundary of the same signature $s$
to the surface, which suggests that the defect rank is conserved.

Consider the case of two radial hedgehogs arranged into a vector field
resembling that of an electric dipole. The hedgehogs must have the
opposite orientations of arrows, as one acts as a source and the other
as a sink.  This structure has $q=0$ by strict topological definition
(Fig.~\ref{fig:fig4}c). However, we observe that the texture on an
enclosing sphere has one positive-signature boundary, which translates
to $q^\ast=2$, the same as we get with na\"ive counting of
hedgehogs (Fig.~\ref{fig:fig4}d). Note that the ``dipole'' in this
case is not the structure with one hyperbolic and one radial hedgehog,
which is usually denoted by this name in nematics. The defect rank
detects this and differentiates between them, while the topological
charge assigns $q=0$ to both.

\begin{figure}
  \includegraphics[width=0.5\textwidth]{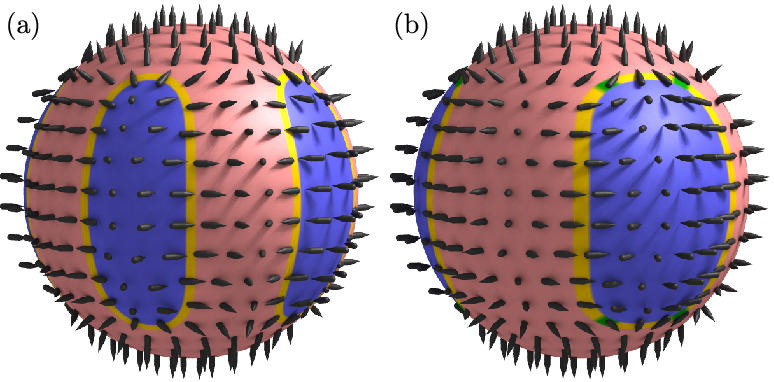}
  \caption{\label{fig:pov}
    Two examples of nematic textures with marked patches (red and blue), boundaries (yellow) and grains (green).
    (a) Analytic model of a nematic point defect with $q=-4$. The
    texture has $5$ patches with $t=-1$, which amounts to $q=-4$. The
    boundaries have signatures $s=-1$, which gives the defect rank of
    $q^\ast=-4$.
    (b) Model of a nematic point defect with $q=4$. There are $3$
    patches with $t=-1$ and $12$ grains with $g=+\thalf$. Together,
    this reproduces $q=4$. However, the defect rank cannot be assigned
    due to the presence of grains.
  }
\end{figure}

The defect rank $q^\ast$ is sign-invariant and matches the existing
conventions \cite{klem} for assigning the charge to the structures in
nematics. Despite similarities, this quantity is not the topological
charge and only stays invariant under restricted set of
transformations that preserve the texture on a chosen surface. The
choice of the surface is no longer arbitrary: different surfaces may
result in different $q^\ast$ even if they enclose the same
topological charge. The difference appears when the contents of one
surface do not fit into the other surface without deviating from the
energy minimum.

The simplicity of texture decomposition and defect rank can be
demonstrated on the model director fields for point defects with known
topological charge \cite{saupe,klem,blaha}. Visualization of the
normal component of the director, $\vec{n}\cdot\boldsymbol{\nu}$,
shows that negative $q$ solutions are realized as $|q|+1$
negative-signature patches on the enclosing sphere
(Fig.~\ref{fig:pov}a). For these solutions, the defect rank can also
be determined and is equal to the topological charge. On the other
hand, positive $q$ solutions have $q-1$ negative-signature patches and
each boundary has four $g=+\thalf$ grains, which gives the correct
topological charge (Fig.~\ref{fig:pov}b), but the defect rank cannot
be determined due to the presence of grains. In nematics, the entire
texture can be arbitrarily multiplied by $-1$, so the $q$ and $-q$
textures are topologically equal. However, the surface textures and
their defect ranks are different, which implies that the two states
are not interchangable without deviating from the energy minimum.
Textures with high defect rank or topological charge are energetically
expensive and as such hard to find in real systems. In an idealized
calculation, the only stable point defects are those with $q=\pm 1$
and in a cluster of many point defects, they prefer to form pairs with
zero total charge \cite{harmonic}. Complex textures are most likely to
be found locally as a part of a larger complex of line defects and
colloidal inclusions, or in frustrated systems.

\section{Conclusion}

Topological defects in directional media are interesting both from the
topological and geometrical point of view. The physical system cannot
undergo a transition between states with different topology, which
stabilizes the defect states. On the other hand, contributions to the
free energy that result from spatial variations of the director depend
solely on the geometry of the field. Constraints conditioned by the
free energy are physically just as important as topological, which
calls for a deeper analysis.

Instead of treating the nematic director field as a separate entity
from the coordinate space, we observed it relatively to a chosen
surface. Our geometric interpretation of topological charge enables
directly relating the geometric appearance of the director field to
its topology. With the texture decomposition, the topological charge
can be found simply by finding the loops where the director is
tangential to the surface and counting them, which is a useful
representation also for the educational purposes. This manner of
calculating the topological charge works also for chiral nematics and
other media with unit vector fields, such as Heisenberg ferromagnets
\cite{kotiuga} and antiferromagnets \cite{dzyalo}.

We show that in nematics, conditions that energetically stabilize the
structures are closely related to the conservation of surface
textures. As the topological charge is ambiguously defined in this
case, we exploit the geometric nature of the texture decomposition and
introduce the defect rank, which measures the shape and structure of
the director field instead of only the topological properties. The
defect rank closely matches the intuitive understanding of defects and
mathematically quantifies the visual clues that are usually used to
identify the defects by eye.  The even or odd parity of the defect
rank also matches the parity of the topological charge, whenever the
defect rank is well-defined. Especially in interpretation of the
experimental data and description of nematic textures, the defect rank
should be used instead of the topological charge. The latter should
not even be assigned a sign when used in a description of a single
defect structure, while the defect rank spans an entire integer range
without ambiguities. This contrasts with the ferromagnetic case, which
is fully described by the topological charge and does not require
description with the defect rank.

Furthermore, the description of defects using surface textures can be
applied to systems with complex boundary conditions, like nematic
colloids \cite{zum4,igor}, faceted particles \cite{facet}, nematic
droplets \cite{dropl} and nematics in porous media \cite{belini}. For
a nematic in a homeotropic micro-channel network, counting of patches
corresponds to the formalism of counting escape directions at channel
crossings, devised by Serra {\sl et al.} \cite{serra}. The confinement
borders of the nematic can be viewed as an enclosing surface and the
decomposition directly related to the boundary conditions enforced by
the surface treatment. Colloidal particles with heterogeneous surface
anchoring also imply a certain texture on its surface
\cite{conradi}. Knowing the meaning of the texture features can
simplify the design of particles with desired topological properties
and aid in interpretation of the results.

\section{Acknowledgments}
The authors acknowledge the support by Slovenian Research Agency under
the contracts P1-0099 \& J1-2335, NAMASTE Center of Excellence, and
HIERARCHY FP7 network 215851-2.


\begin{thebibliography}{28}%
\makeatletter
\providecommand \@ifxundefined [1]{%
 \@ifx{#1\undefined}
}%
\providecommand \@ifnum [1]{%
 \ifnum #1\expandafter \@firstoftwo
 \else \expandafter \@secondoftwo
 \fi
}%
\providecommand \@ifx [1]{%
 \ifx #1\expandafter \@firstoftwo
 \else \expandafter \@secondoftwo
 \fi
}%
\providecommand \natexlab [1]{#1}%
\providecommand \enquote  [1]{``#1''}%
\providecommand \bibnamefont  [1]{#1}%
\providecommand \bibfnamefont [1]{#1}%
\providecommand \citenamefont [1]{#1}%
\providecommand \href@noop [0]{\@secondoftwo}%
\providecommand \href [0]{\begingroup \@sanitize@url \@href}%
\providecommand \@href[1]{\@@startlink{#1}\@@href}%
\providecommand \@@href[1]{\endgroup#1\@@endlink}%
\providecommand \@sanitize@url [0]{\catcode `\\12\catcode `\$12\catcode
  `\&12\catcode `\#12\catcode `\^12\catcode `\_12\catcode `\%12\relax}%
\providecommand \@@startlink[1]{}%
\providecommand \@@endlink[0]{}%
\providecommand \url  [0]{\begingroup\@sanitize@url \@url }%
\providecommand \@url [1]{\endgroup\@href {#1}{\urlprefix }}%
\providecommand \urlprefix  [0]{URL }%
\providecommand \Eprint [0]{\href }%
\providecommand \doibase [0]{http://dx.doi.org/}%
\providecommand \selectlanguage [0]{\@gobble}%
\providecommand \bibinfo  [0]{\@secondoftwo}%
\providecommand \bibfield  [0]{\@secondoftwo}%
\providecommand \translation [1]{[#1]}%
\providecommand \BibitemOpen [0]{}%
\providecommand \bibitemStop [0]{}%
\providecommand \bibitemNoStop [0]{.\EOS\space}%
\providecommand \EOS [0]{\spacefactor3000\relax}%
\providecommand \BibitemShut  [1]{\csname bibitem#1\endcsname}%
\let\auto@bib@innerbib\@empty
\bibitem [{\citenamefont {Mu\v{s}evi\v{c}}\ \emph {et~al.}(2006)\citenamefont
  {Mu\v{s}evi\v{c}}, \citenamefont {\v{S}karabot}, \citenamefont {Tkalec}, \citenamefont
  {Ravnik},\ and\ \citenamefont {\v{Z}umer}}]{musevic}%
  \BibitemOpen
  \bibfield  {author} {\bibinfo {author} {\bibfnamefont {I.}~\bibnamefont
  {Mu\v{s}evi\v{c}}}, \bibinfo {author} {\bibfnamefont {M.}~\bibnamefont
  {\v{S}karabot}}, \bibinfo {author} {\bibfnamefont {U.}~\bibnamefont {Tkalec}},
  \bibinfo {author} {\bibfnamefont {M.}~\bibnamefont {Ravnik}}, \ and\ \bibinfo
  {author} {\bibfnamefont {S.}~\bibnamefont {\v{Z}umer}},\ }\href@noop {}
  {\bibfield  {journal} {\bibinfo  {journal} {Science}\ }\textbf {\bibinfo
  {volume} {18}},\ \bibinfo {pages} {954} (\bibinfo {year} {2006})}\BibitemShut
  {NoStop}%
\bibitem [{\citenamefont {Araki}\ \emph {et~al.}(2011)\citenamefont {Araki},
  \citenamefont {Buscaglia}, \citenamefont {Bellini},\ and\ \citenamefont
  {Tanaka}}]{belini}%
  \BibitemOpen
  \bibfield  {author} {\bibinfo {author} {\bibfnamefont {T.}~\bibnamefont
  {Araki}}, \bibinfo {author} {\bibfnamefont {M.}~\bibnamefont {Buscaglia}},
  \bibinfo {author} {\bibfnamefont {T.}~\bibnamefont {Bellini}}, \ and\
  \bibinfo {author} {\bibfnamefont {H.}~\bibnamefont {Tanaka}},\ }\href@noop {}
  {\bibfield  {journal} {\bibinfo  {journal} {Nat.~Mat.}\ }\textbf {\bibinfo
  {volume} {10}},\ \bibinfo {pages} {303} (\bibinfo {year} {2011})}\BibitemShut
  {NoStop}%
\bibitem [{\citenamefont {Serra}\ \emph {et~al.}(2011)\citenamefont {Serra},
  \citenamefont {Vishnubhatla}, \citenamefont {Buscaglia}, \citenamefont
  {Cerbino}, \citenamefont {Osellame}, \citenamefont {Cerullo},\ and\
  \citenamefont {Bellini}}]{serra}%
  \BibitemOpen
  \bibfield  {author} {\bibinfo {author} {\bibfnamefont {F.}~\bibnamefont
  {Serra}}, \bibinfo {author} {\bibfnamefont {K.~C.}\ \bibnamefont
  {Vishnubhatla}}, \bibinfo {author} {\bibfnamefont {M.}~\bibnamefont
  {Buscaglia}}, \bibinfo {author} {\bibfnamefont {R.}~\bibnamefont {Cerbino}},
  \bibinfo {author} {\bibfnamefont {R.}~\bibnamefont {Osellame}}, \bibinfo
  {author} {\bibfnamefont {G.}~\bibnamefont {Cerullo}}, \ and\ \bibinfo
  {author} {\bibfnamefont {T.}~\bibnamefont {Bellini}},\ }\href@noop {}
  {\bibfield  {journal} {\bibinfo  {journal} {Soft Matter}\ }\textbf {\bibinfo
  {volume} {7}},\ \bibinfo {pages} {10945} (\bibinfo {year}
  {2011})}\BibitemShut {NoStop}%
\bibitem [{\citenamefont {Humar}\ \emph {et~al.}(2009)\citenamefont {Humar},
  \citenamefont {Ravnik}, \citenamefont {Pajk},\ and\ \citenamefont
  {Mu\v{s}evi\v{c}}}]{humar}%
  \BibitemOpen
  \bibfield  {author} {\bibinfo {author} {\bibfnamefont {M.}~\bibnamefont
  {Humar}}, \bibinfo {author} {\bibfnamefont {M.}~\bibnamefont {Ravnik}},
  \bibinfo {author} {\bibfnamefont {S.}~\bibnamefont {Pajk}}, \ and\ \bibinfo
  {author} {\bibfnamefont {I.}~\bibnamefont {Mu\v{s}evi\v{c}}},\ }\href@noop {}
  {\bibfield  {journal} {\bibinfo  {journal} {Nat. Photonics}\ }\textbf
  {\bibinfo {volume} {3}},\ \bibinfo {pages} {595} (\bibinfo {year}
  {2009})}\BibitemShut {NoStop}%
\bibitem [{\citenamefont {de~Gennes}\ and\ \citenamefont
  {Prost}(1993)}]{degennes}%
  \BibitemOpen
  \bibfield  {author} {\bibinfo {author} {\bibfnamefont {P.~G.}\ \bibnamefont
  {de~Gennes}}\ and\ \bibinfo {author} {\bibfnamefont {J.}~\bibnamefont
  {Prost}},\ }\href@noop {} {\emph {\bibinfo {title} {The physics of liquid
  crystals}}}\ (\bibinfo  {publisher} {Oxford University Press},\ \bibinfo
  {year} {1993})\BibitemShut {NoStop}%
\bibitem [{\citenamefont {Mermin}(1979)}]{mermin}%
  \BibitemOpen
  \bibfield  {author} {\bibinfo {author} {\bibfnamefont {N.~D.}\ \bibnamefont
  {Mermin}},\ }\href@noop {} {\bibfield  {journal} {\bibinfo  {journal}
  {Rev.~Mod.~Phys.}\ }\textbf {\bibinfo {volume} {51}},\ \bibinfo {pages} {591}
  (\bibinfo {year} {1979})}\BibitemShut {NoStop}%
\bibitem [{\citenamefont {Trebin}(1982)}]{trebinrev}%
  \BibitemOpen
  \bibfield  {author} {\bibinfo {author} {\bibfnamefont {H.~R.}\ \bibnamefont
  {Trebin}},\ }\href@noop {} {\bibfield  {journal} {\bibinfo  {journal}
  {Adv.~Phys.}\ }\textbf {\bibinfo {volume} {31}},\ \bibinfo {pages} {195}
  (\bibinfo {year} {1982})}\BibitemShut {NoStop}%
\bibitem [{\citenamefont {Volovik}\ and\ \citenamefont
  {Mineev}(1986)}]{volomin}%
  \BibitemOpen
  \bibfield  {author} {\bibinfo {author} {\bibfnamefont {G.~E.}\ \bibnamefont
  {Volovik}}\ and\ \bibinfo {author} {\bibfnamefont {V.~P.}\ \bibnamefont
  {Mineev}},\ }\href@noop {} {\bibfield  {journal} {\bibinfo  {journal}
  {Sov.~Phys.~JETP}\ }\textbf {\bibinfo {volume} {45}},\ \bibinfo {pages}
  {1186} (\bibinfo {year} {1986})}\BibitemShut {NoStop}%
\bibitem [{\citenamefont {Kleman}(1973)}]{klemanrev}%
  \BibitemOpen
  \bibfield  {author} {\bibinfo {author} {\bibfnamefont {M.}~\bibnamefont
  {Kleman}},\ }\href@noop {} {\bibfield  {journal} {\bibinfo  {journal}
  {Phil.~Mag.}\ }\textbf {\bibinfo {volume} {27}},\ \bibinfo {pages} {1057}
  (\bibinfo {year} {1973})}\BibitemShut {NoStop}%
\bibitem [{\citenamefont {Kleman}\ and\ \citenamefont
  {Lavrentovich}(2006)}]{klem}%
  \BibitemOpen
  \bibfield  {author} {\bibinfo {author} {\bibfnamefont {M.}~\bibnamefont
  {Kleman}}\ and\ \bibinfo {author} {\bibfnamefont {O.~D.}\ \bibnamefont
  {Lavrentovich}},\ }\href@noop {} {\bibfield  {journal} {\bibinfo  {journal}
  {Phil.~Mag.}\ }\textbf {\bibinfo {volume} {86}},\ \bibinfo {pages} {4117}
  (\bibinfo {year} {2006})}\BibitemShut {NoStop}%
\bibitem [{\citenamefont {Kotiuga}(1989)}]{kotiuga}%
  \BibitemOpen
  \bibfield  {author} {\bibinfo {author} {\bibfnamefont {P.~R.}\ \bibnamefont
  {Kotiuga}},\ }\href@noop {} {\bibfield  {journal} {\bibinfo  {journal} {IEEE
  Transactions on magnetics}\ }\textbf {\bibinfo {volume} {25}},\ \bibinfo
  {pages} {3476} (\bibinfo {year} {1989})}\BibitemShut {NoStop}%
\bibitem [{\citenamefont {Kamien}(2002)}]{geom1}%
  \BibitemOpen
  \bibfield  {author} {\bibinfo {author} {\bibfnamefont {R.~D.}\ \bibnamefont
  {Kamien}},\ }\href@noop {} {\bibfield  {journal} {\bibinfo  {journal}
  {Rev.~Mod.~Phys.}\ }\textbf {\bibinfo {volume} {74}} (\bibinfo {year}
  {2002})}\BibitemShut {NoStop}%
\bibitem [{\citenamefont {Lavrentovich}(1998)}]{lavworlds}%
  \BibitemOpen
  \bibfield  {author} {\bibinfo {author} {\bibfnamefont {O.~D.}\ \bibnamefont
  {Lavrentovich}},\ }\href@noop {} {\bibfield  {journal} {\bibinfo  {journal}
  {Liq.~Cryst.}\ }\textbf {\bibinfo {volume} {24}},\ \bibinfo {pages} {117}
  (\bibinfo {year} {1998})}\BibitemShut {NoStop}%
\bibitem [{\citenamefont {Kurik}\ and\ \citenamefont
  {Lavrentovich}(1988)}]{kurik}%
  \BibitemOpen
  \bibfield  {author} {\bibinfo {author} {\bibfnamefont {M.~V.}\ \bibnamefont
  {Kurik}}\ and\ \bibinfo {author} {\bibfnamefont {O.~D.}\ \bibnamefont
  {Lavrentovich}},\ }\href@noop {} {\bibfield  {journal} {\bibinfo  {journal}
  {Sov.~Phys.~Usp.}\ }\textbf {\bibinfo {volume} {31}},\ \bibinfo {pages} {196}
  (\bibinfo {year} {1988})}\BibitemShut {NoStop}%
\bibitem [{\citenamefont {Volovik}\ and\ \citenamefont
  {Lavrentovich}(1984)}]{volovik}%
  \BibitemOpen
  \bibfield  {author} {\bibinfo {author} {\bibfnamefont {G.~E.}\ \bibnamefont
  {Volovik}}\ and\ \bibinfo {author} {\bibfnamefont {O.~D.}\ \bibnamefont
  {Lavrentovich}},\ }\href@noop {} {\bibfield  {journal} {\bibinfo  {journal}
  {Sov.~Phys.~JETP}\ }\textbf {\bibinfo {volume} {58}},\ \bibinfo {pages}
  {1159} (\bibinfo {year} {1984})}\BibitemShut {NoStop}%
\bibitem [{\citenamefont {Kleman}(1983)}]{walls}%
  \BibitemOpen
  \bibfield  {author} {\bibinfo {author} {\bibfnamefont {M.}~\bibnamefont
  {Kleman}},\ }\href@noop {} {\emph {\bibinfo {title} {Points, Lines and
  Walls}}}\ (\bibinfo  {publisher} {John Wiley \& Sons Inc.},\ \bibinfo {year}
  {1983})\BibitemShut {NoStop}%
\bibitem [{\citenamefont {Nabarro}(1972)}]{nabarro}%
  \BibitemOpen
  \bibfield  {author} {\bibinfo {author} {\bibfnamefont {F.~R.~N.}\
  \bibnamefont {Nabarro}},\ }\href@noop {} {\bibfield  {journal} {\bibinfo
  {journal} {J.~Phys.}\ }\textbf {\bibinfo {volume} {33}},\ \bibinfo {pages}
  {1089} (\bibinfo {year} {1972})}\BibitemShut {NoStop}%
\bibitem [{\citenamefont {J\"anich}(1987)}]{janich}%
  \BibitemOpen
  \bibfield  {author} {\bibinfo {author} {\bibfnamefont {K.}~\bibnamefont
  {J\"anich}},\ }\href@noop {} {\bibfield  {journal} {\bibinfo  {journal}
  {Acta~Appl.~Math.}\ }\textbf {\bibinfo {volume} {8}},\ \bibinfo {pages} {65}
  (\bibinfo {year} {1987})}\BibitemShut {NoStop}%
\bibitem [{\citenamefont {Lavrentovich}\ and\ \citenamefont
  {Terent'ev}(1986)}]{terent}%
  \BibitemOpen
  \bibfield  {author} {\bibinfo {author} {\bibfnamefont {O.~D.}\ \bibnamefont
  {Lavrentovich}}\ and\ \bibinfo {author} {\bibfnamefont {E.~M.}\ \bibnamefont
  {Terent'ev}},\ }\href@noop {} {\bibfield  {journal} {\bibinfo  {journal}
  {Sov.~Phys.~JETP}\ }\textbf {\bibinfo {volume} {64}},\ \bibinfo {pages}
  {1237} (\bibinfo {year} {1986})}\BibitemShut {NoStop}%
\bibitem [{\citenamefont {Saupe}(1973)}]{saupe}%
  \BibitemOpen
  \bibfield  {author} {\bibinfo {author} {\bibfnamefont {A.}~\bibnamefont
  {Saupe}},\ }\href@noop {} {\bibfield  {journal} {\bibinfo  {journal}
  {Mol.~Cryst.~Liq.~Cryst.}\ }\textbf {\bibinfo {volume} {21}},\ \bibinfo
  {pages} {211} (\bibinfo {year} {1973})}\BibitemShut {NoStop}%
\bibitem [{\citenamefont {Blaha}(1976)}]{blaha}%
  \BibitemOpen
  \bibfield  {author} {\bibinfo {author} {\bibfnamefont {S.}~\bibnamefont
  {Blaha}},\ }\href@noop {} {\bibfield  {journal} {\bibinfo  {journal}
  {Phys.~Rev.~Lett.}\ }\textbf {\bibinfo {volume} {36}},\ \bibinfo {pages}
  {874} (\bibinfo {year} {1976})}\BibitemShut {NoStop}%
\bibitem [{\citenamefont {Brezis}\ \emph {et~al.}(1986)\citenamefont {Brezis},
  \citenamefont {Coron},\ and\ \citenamefont {Lieb}}]{harmonic}%
  \BibitemOpen
  \bibfield  {author} {\bibinfo {author} {\bibfnamefont {H.}~\bibnamefont
  {Brezis}}, \bibinfo {author} {\bibfnamefont {J.}~\bibnamefont {Coron}}, \
  and\ \bibinfo {author} {\bibfnamefont {E.~H.}\ \bibnamefont {Lieb}},\
  }\href@noop {} {\bibfield  {journal} {\bibinfo  {journal}
  {Commmun.~Math.~Phys.}\ }\textbf {\bibinfo {volume} {107}},\ \bibinfo {pages}
  {649} (\bibinfo {year} {1986})}\BibitemShut {NoStop}%
\bibitem [{\citenamefont {Dzyaloshkinskii}(1977)}]{dzyalo}%
  \BibitemOpen
  \bibfield  {author} {\bibinfo {author} {\bibfnamefont {I.~E.}\ \bibnamefont
  {Dzyaloshkinskii}},\ }\href@noop {} {\bibfield  {journal} {\bibinfo
  {journal} {JETP.~Lett.}\ }\textbf {\bibinfo {volume} {25}},\ \bibinfo {pages}
  {98} (\bibinfo {year} {1977})}\BibitemShut {NoStop}%
\bibitem [{\citenamefont {Ravnik}\ and\ \citenamefont {\v{Z}umer}(2009)}]{zum4}%
  \BibitemOpen
  \bibfield  {author} {\bibinfo {author} {\bibfnamefont {M.}~\bibnamefont
  {Ravnik}}\ and\ \bibinfo {author} {\bibfnamefont {S.}~\bibnamefont
  {\v{Z}umer}},\ }\href@noop {} {\bibfield  {journal} {\bibinfo  {journal} {Soft
  Matter}\ }\textbf {\bibinfo {volume} {5}},\ \bibinfo {pages} {4520} (\bibinfo
  {year} {2009})}\BibitemShut {NoStop}%
\bibitem [{\citenamefont {Tkalec}\ \emph {et~al.}(2011)\citenamefont {Tkalec},
  \citenamefont {Ravnik}, \citenamefont {\v{C}opar}, \citenamefont {\v{Z}umer},\ and\
  \citenamefont {Mu\v{s}evi\v{c}}}]{igor}%
  \BibitemOpen
  \bibfield  {author} {\bibinfo {author} {\bibfnamefont {U.}~\bibnamefont
  {Tkalec}}, \bibinfo {author} {\bibfnamefont {M.}~\bibnamefont {Ravnik}},
  \bibinfo {author} {\bibfnamefont {S.}~\bibnamefont {\v{C}opar}}, \bibinfo
  {author} {\bibfnamefont {S.}~\bibnamefont {\v{Z}umer}}, \ and\ \bibinfo {author}
  {\bibfnamefont {I.}~\bibnamefont {Mu\v{s}evi\v{c}}},\ }\href@noop {} {\bibfield
  {journal} {\bibinfo  {journal} {Science}\ }\textbf {\bibinfo {volume}
  {333}},\ \bibinfo {pages} {64} (\bibinfo {year} {2011})}\BibitemShut
  {NoStop}%
\bibitem [{\citenamefont {Phillips}\ and\ \citenamefont {Rey}(2011)}]{facet}%
  \BibitemOpen
  \bibfield  {author} {\bibinfo {author} {\bibfnamefont {P.~M.}\ \bibnamefont
  {Phillips}}\ and\ \bibinfo {author} {\bibfnamefont {A.~D.}\ \bibnamefont
  {Rey}},\ }\href@noop {} {\bibfield  {journal} {\bibinfo  {journal} {Soft
  Matter}\ }\textbf {\bibinfo {volume} {7}},\ \bibinfo {pages} {2052} (\bibinfo
  {year} {2011})}\BibitemShut {NoStop}%
\bibitem [{\citenamefont {Prishchepa}\ \emph {et~al.}(2005)\citenamefont
  {Prischepa}, \citenamefont {Shabanov},\ and\ \citenamefont
  {Zyryanov}}]{dropl}%
  \BibitemOpen
  \bibfield  {author} {\bibinfo {author} {\bibfnamefont {O.~O.}\ \bibnamefont
  {Prischepa}}, \bibinfo {author} {\bibfnamefont {A.~V.}\ \bibnamefont
  {Shabanov}}, \ and\ \bibinfo {author} {\bibfnamefont {V.~Y.}\ \bibnamefont
  {Zyryanov}},\ }\href@noop {} {\bibfield  {journal} {\bibinfo  {journal}
  {Phys.~Rev.~E}\ }\textbf {\bibinfo {volume} {72}},\ \bibinfo {pages} {031712}
  (\bibinfo {year} {2005})}\BibitemShut {NoStop}%
\bibitem [{\citenamefont {Conradi}\ \emph {et~al.}(2010)\citenamefont
  {Conradi}, \citenamefont {Zorko},\ and\ \citenamefont
  {Mu\v{s}evi\v{c}}}]{conradi}%
  \BibitemOpen
  \bibfield  {author} {\bibinfo {author} {\bibfnamefont {M.}~\bibnamefont
  {Conradi}}, \bibinfo {author} {\bibfnamefont {M.}~\bibnamefont {Zorko}}, \
  and\ \bibinfo {author} {\bibfnamefont {I.}~\bibnamefont {Mu\v{s}evi\v{c}}},\
  }\href@noop {} {\bibfield  {journal} {\bibinfo  {journal} {Opt.~express}\
  }\textbf {\bibinfo {volume} {18}},\ \bibinfo {pages} {500} (\bibinfo {year}
  {2010})}\BibitemShut {NoStop}%
\end{thebibliography}
\end{document}